%% file: main.tex
\documentclass[10pt,twocolumn,letterpaper]{article}

\usepackage{cvpr}              
\usepackage{graphicx}
\input{preamble}

\definecolor{cvprblue}{rgb}{0.21,0.49,0.74}
\usepackage[pagebackref,breaklinks,colorlinks,citecolor=cvprblue]{hyperref}

\def\paperID{*****} 
\def\confName{CVPR}
\def\confYear{2024}

\title{{Custom Cloth Creation and Virtual Try-on for Everyone}}

\author{
\quad Pei Chen\textsuperscript{\rm 1}
\quad Heng Wang\textsuperscript{\rm 1}
\quad Sainan Sun\textsuperscript{\rm 1}
\quad Zhiyuan Chen\textsuperscript{\rm 1} \\
\quad Zhenkun Liu\textsuperscript{\rm 1}
\quad Shuhua Cao\textsuperscript{\rm 1}
\quad Li Yang\textsuperscript{\rm 1}
\quad Minghui Yang\textsuperscript{\rm 1}\thanks{Corresponding author} \\
\textsuperscript{\rm 1}Ant Group\\
{\tt\small \{azhe.cp, feixi.wh, ssn01369727, juzhen.czy, liuzhenkun.lzk, caoshuhua.csh\}@antgroup.com,} \\ {\tt\small yl406239@digital-engine.com, minghui.ymh@antgroup.com}
}

\begin{document}
\maketitle
\input{sec/0_abstract}    
\input{sec/1_intro}
\input{sec/2_method}
\input{sec/3_demonstration}
\input{sec/4_conclusion}


\input{main.bbl}


\end{document}

%% file: preamble.tex
\usepackage[dvipsnames]{xcolor}


%% file: sec/0_abstract.tex
\begin{abstract}
This demo showcases a simple tool that utilizes AIGC technology, enabling both professional designers and regular users to easily customize clothing for their digital avatars. Customization options include changing clothing colors, textures, logos, and patterns. Compared with traditional 3D modeling processes, our approach significantly enhances efficiency and interactivity and reduces production costs.
\end{abstract}

%% file: sec/1_intro.tex
\section{Introduction}

In recent years, virtual avatars have emerged as a significant platform for showcasing the external image of various technological products, due to their rich expressiveness. As a critical element of digital representation, there is a substantial business demand for the creation of virtual clothing to address diverse scenarios such as seasonal changes, promotional activities, and holiday celebrations. However, designing, modeling, and binding 3D assets such as digital clothing and accessories is a complex and time-consuming process. Typically, the procurement of 3D assets is costly and time-consuming, often failing to meet immediate business needs. This necessitates algorithmic improvements to enhance efficiency. \\

\begin{figure}
  \centering
  \includegraphics[width=0.48\linewidth]{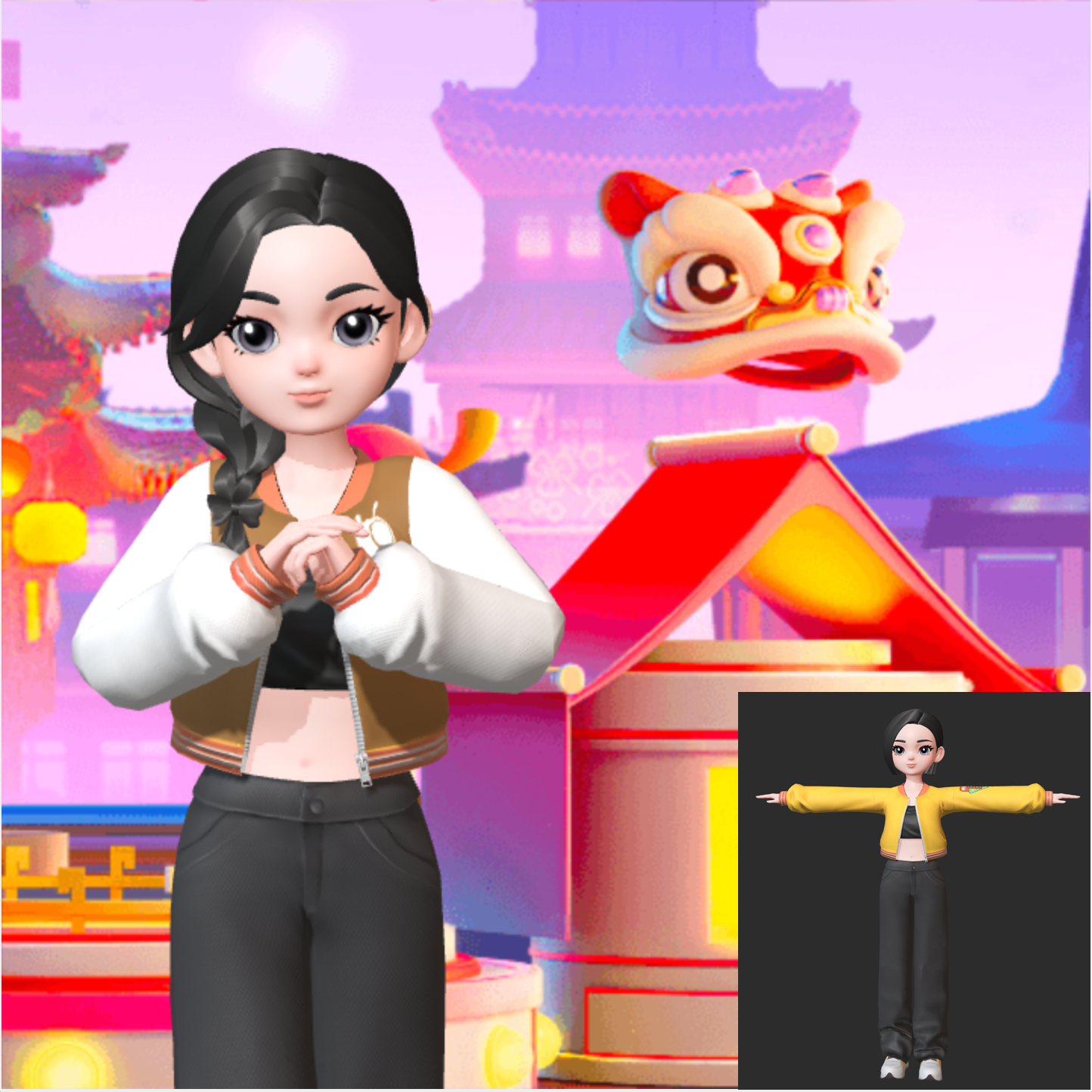}
  \includegraphics[width=0.48\linewidth]{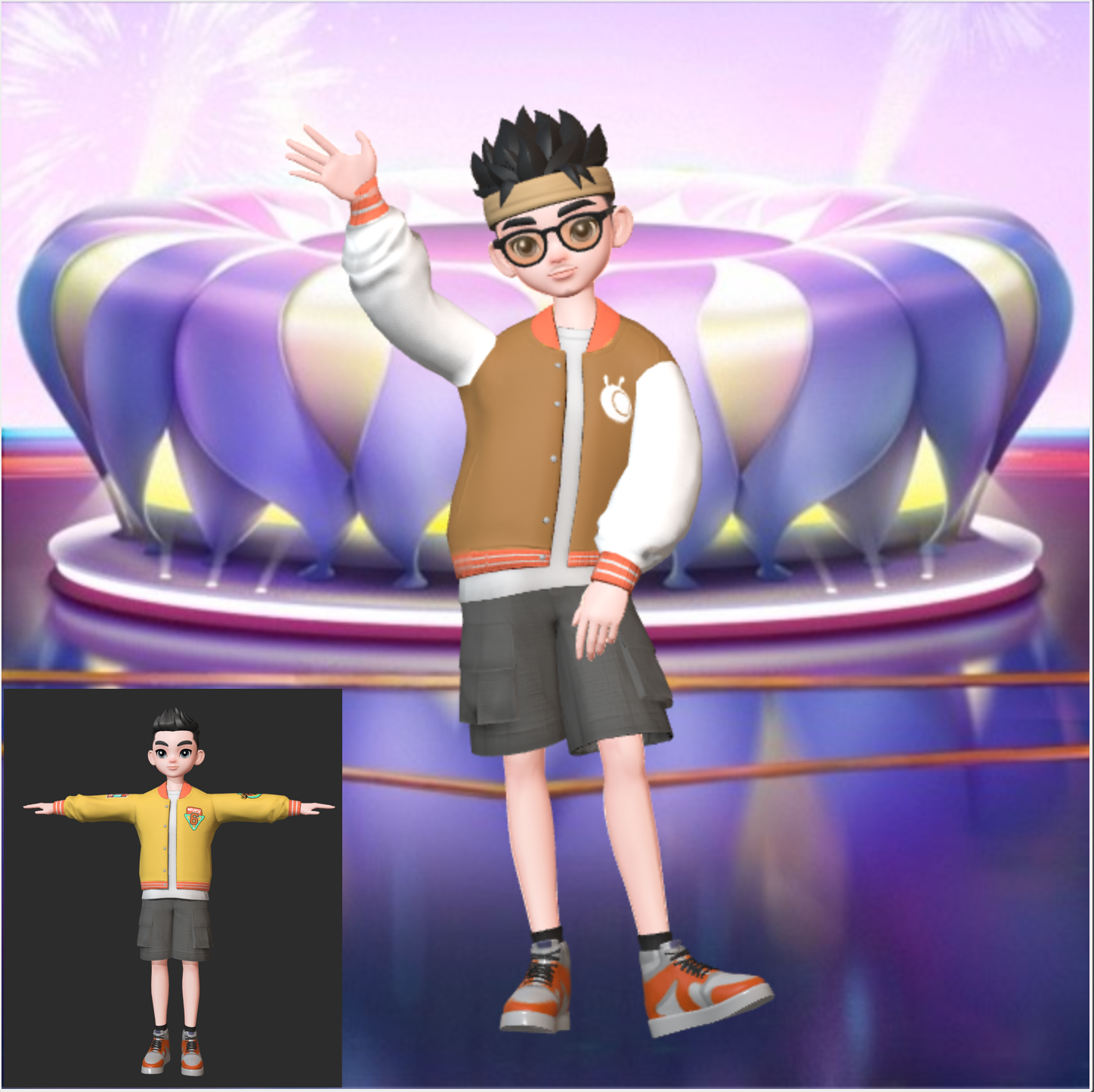}
  \caption{\textbf{Texture Editing Example}. The main image shows the clothing after texture editing, with the original clothing shown in the bottom right or bottom left corner.}
\label{fig:examples}
\end{figure}

In the field of 3D modeling, thanks to recent developments, individuals can now readily acquire prefabricated 3D asset materials from various online platforms, encompassing items like long/short sleeves, pants/shorts, skirts, jackets, and coats. We recognize that by leveraging the popular AI-generated content (AIGC) technology, it is feasible to swiftly and economically customize these existing assets by altering their surface textures, as shown in Fig.~\ref{fig:examples}. This approach facilitates rapid and cost-effective customization of 3D clothing, thereby enabling diversified and personalized editing of 3D assets.

Additionally, this method of texture editing for 3D clothing can also be applied to user-generated content (UGC) products, enabling ordinary users to quickly create their unique digital avatars. This capability enriches the metaverse with more diverse and enjoyable experiences.

%% file: sec/2_method.tex
\section{Method}

\begin{figure*}[ht]
\begin{center}
  \includegraphics[width=1.0\linewidth]{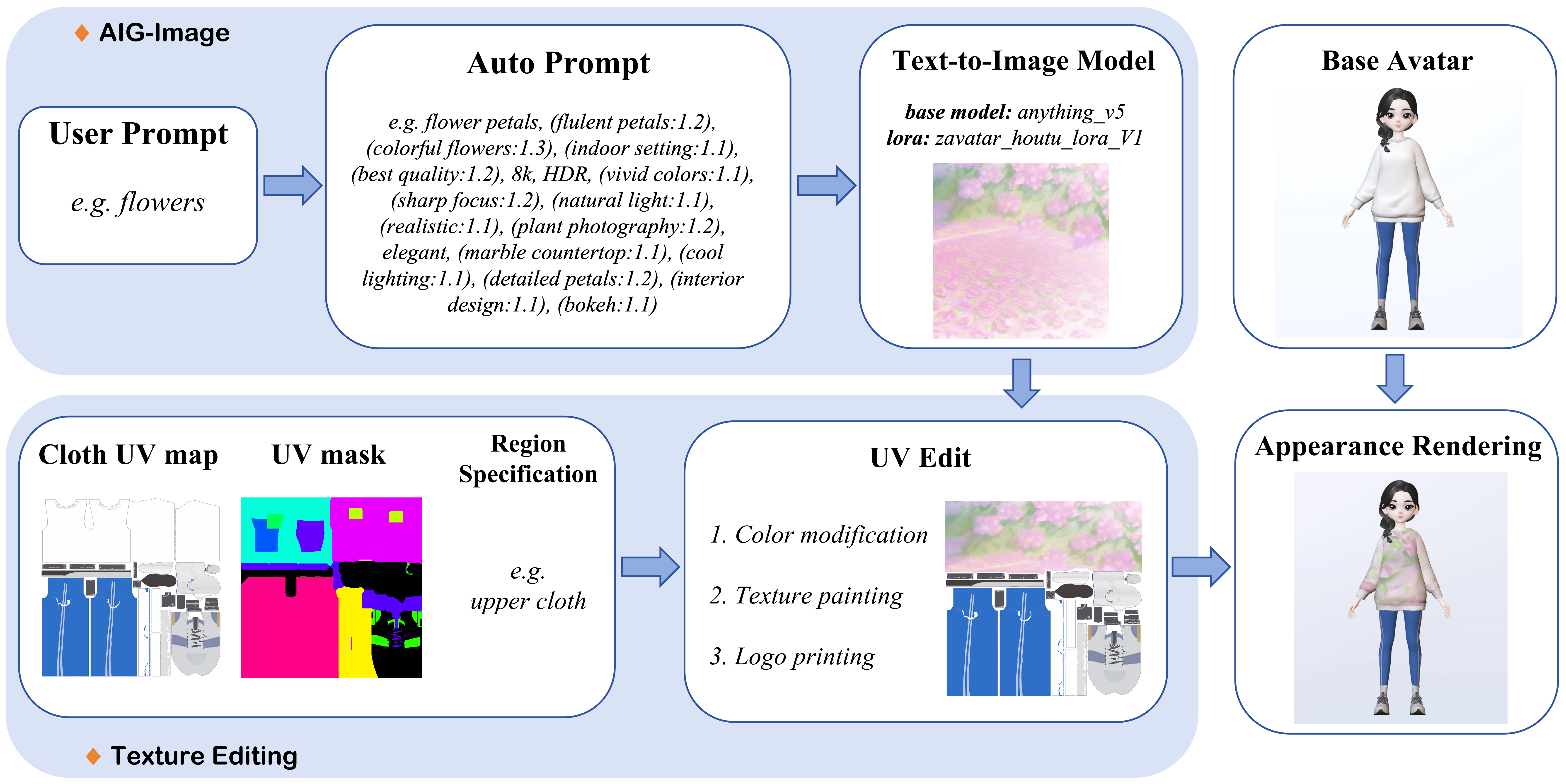}
\end{center}
  \caption{\textbf{Three Modes of Clothing Texture Editing.} Here are three editing modes that meet different types of custom clothing requirements: localized color or texture modifications, logo printing.}
\label{fig:pipeline}
\end{figure*}

In this section, we will introduce how clothing texture editing is accomplished. Essentially, clothing texture editing is achieved by modifying the UV maps of the garments. During the editing process, users typically prefer to modify specific parts of the clothing rather than completely changing the entire outfit. To achieve this functionality, we first have 3D artists annotate different parts of the garment. Next, to enable ordinary users to easily "draw" their ideas onto clothing, we utilize a text-to-image generation model to generate images. We then provide three modes of clothing texture modification to accomplish the editing process. The overall process is depicted in Fig.~\ref{fig:pipeline} \\

\begin{figure*}[ht]
\begin{center}
  \includegraphics[width=0.95\linewidth]{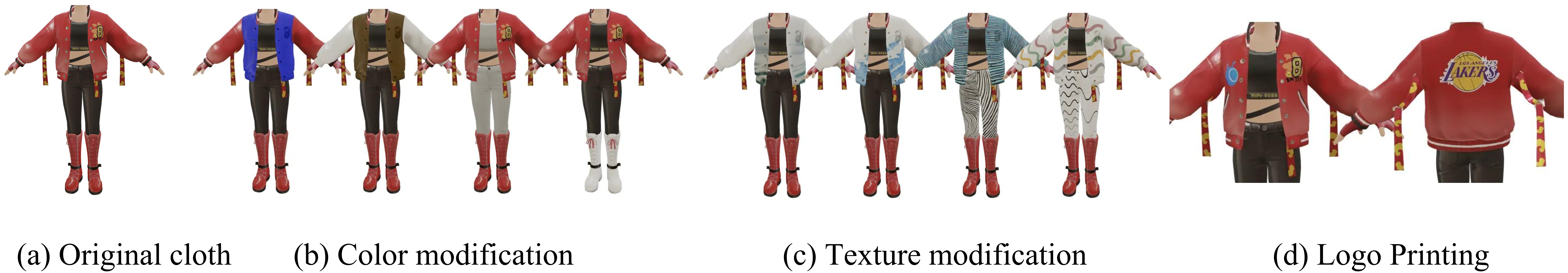}
\end{center}
  \caption{\textbf{Three Modes of Clothing Texture Editing.} Here are three editing modes that meet different types of custom clothing requirements: localized color or texture modifications, logo printing.}
\label{fig:3modes}
\vspace{-5mm}
\end{figure*}

\subsection{Clothing Parts Annotation}

\begin{figure}
  \centering
  \includegraphics[width=0.7\linewidth]{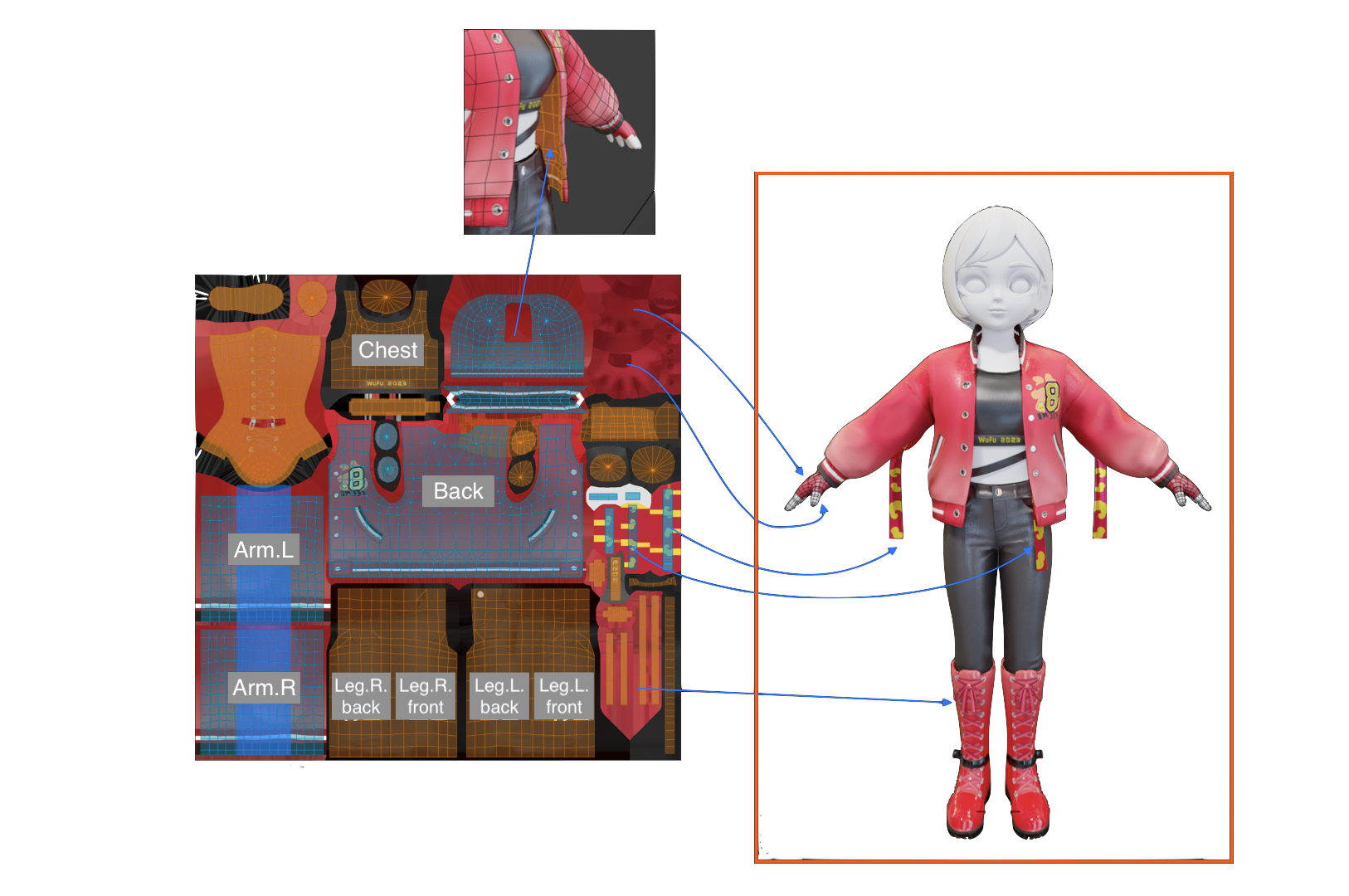}
  \centering
  \includegraphics[width=0.28\linewidth]{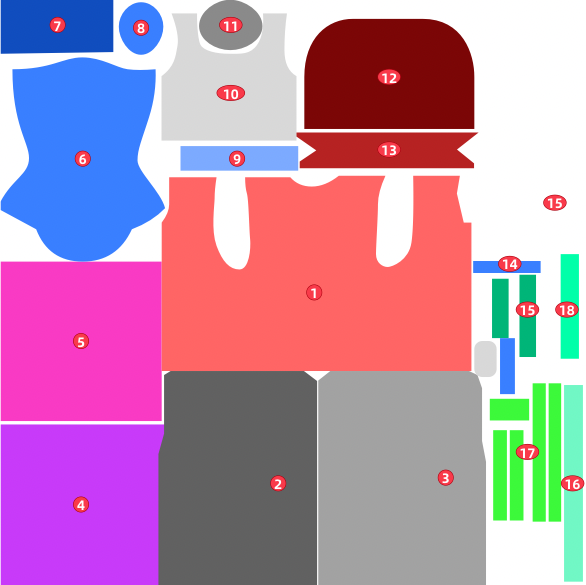}
  \caption{Professional artists use 3D design software to annotate different parts of the clothing manually and generate corresponding mask images of the clothing's UV maps.}
\label{fig:annotation}
\end{figure}

Pre-labeling different parts of 3D clothing allows generated images to be accurately printed in specified locations. 3D artists use \textit{Adobe Substance Painter} software, where they can simply click to select a particular clothing part, and the software automatically highlights the corresponding area on the UV map. Artists can then paint this area with a specific color for labeling, making the process very convenient, as shown in Fig.~\ref{fig:annotation}. With this labeling, when users perform clothing texture editing, the program can retrieve the corresponding clothing parts based on the color labels, facilitating the editing process.

\subsection{Text-to-Image Generation Model}

Text-to-image generation model~\cite{rombach2022high} is the key to achieving customization. To ensure that the generated images are suitable for use as clothing textures, we have specifically trained a base model and several lora models, primarily for generating images in specific styles such as cartoon, aesthetic, or scenic styles.

\subsection{Clothing Texture Editing}

We offer three clothing texture editing modes, as illustrated in Fig.~\ref{fig:3modes}. Color modification involves changing the color of specific parts of the garment, while texture modification entails applying images generated by AIGC to designated areas of the clothing. Logo printing involves using AI-generated images as logos printed onto the clothing.

%% file: sec/3_demonstration.tex
\section{Demonstraion}

\begin{figure}
  \centering
  \includegraphics[width=0.45\linewidth]{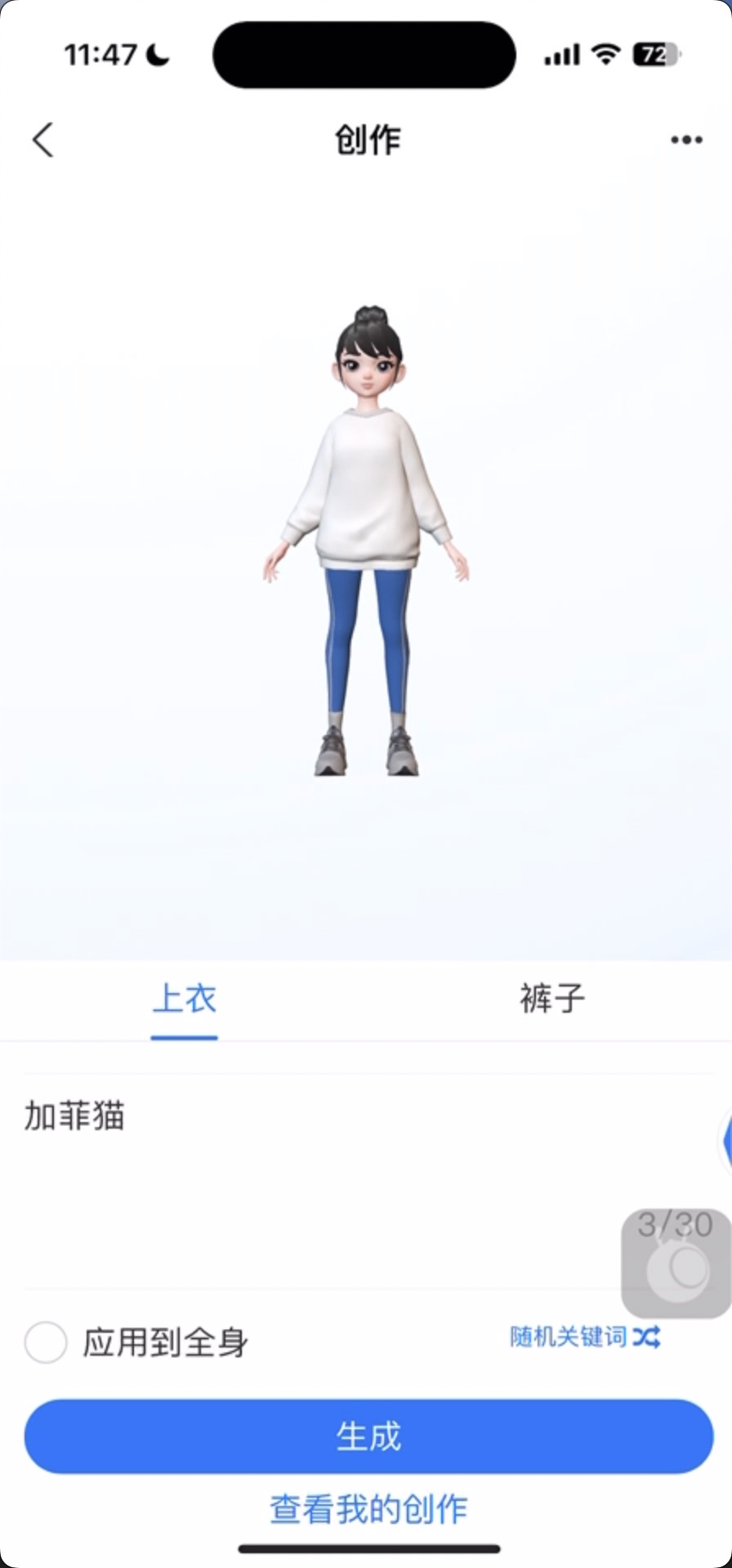}
  \centering
  \includegraphics[width=0.45\linewidth]{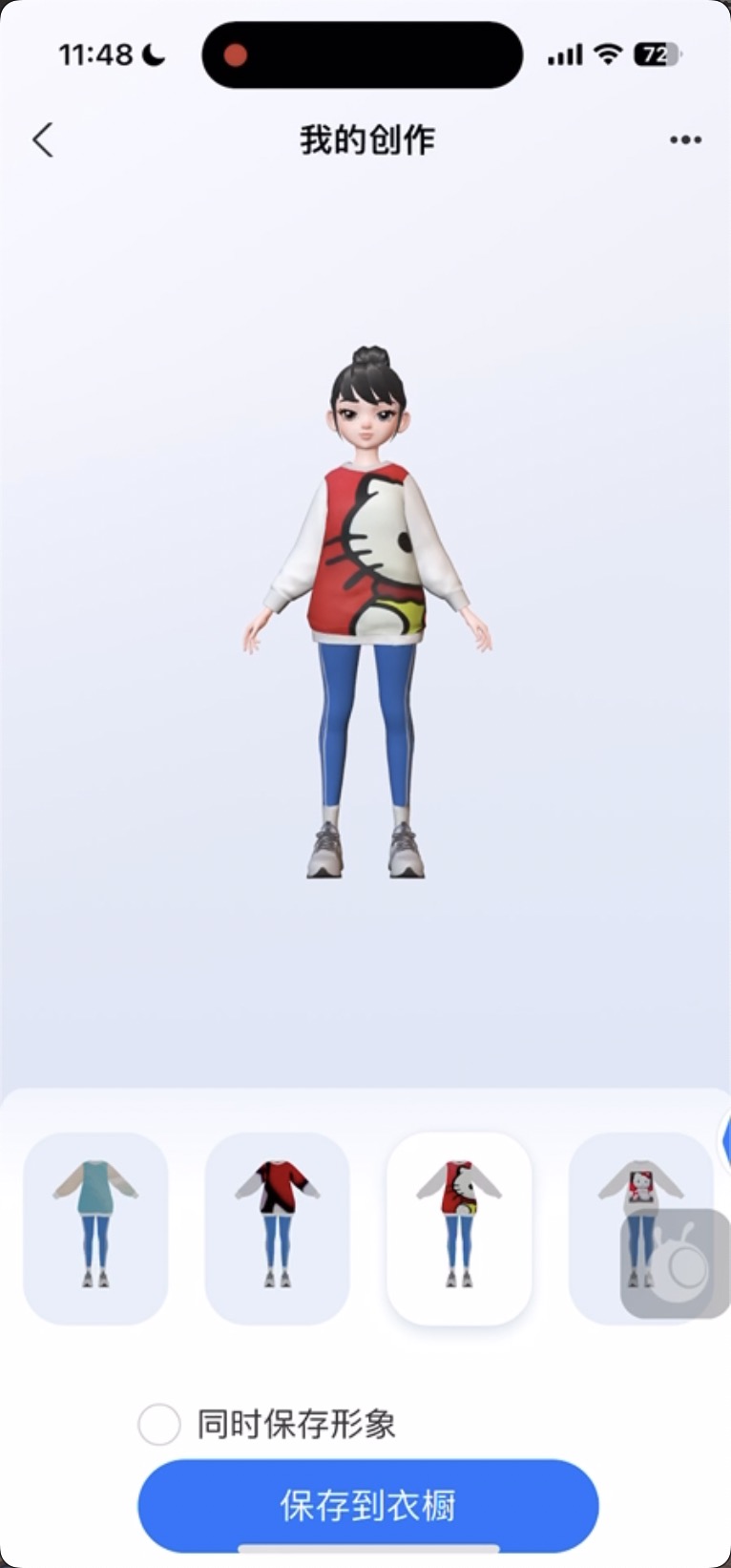}
  \caption{\textbf{Customized Clothing Creation on Alipay app}. Users can use our tool within the Alipay app to create clothing for their digital avatars.}
\label{fig:demo}
\end{figure}

We have also developed a page within the Alipay app where ordinary users can customize their digital clothing. As shown in Fig.~\ref{fig:demo}, users select the specific clothing parts they want to edit and enter prompts in the text box. This process allows them to receive a personalized set of clothing, which can be saved in their virtual wardrobe. This way, users can create their unique digital persona directly within Alipay.

%% file: sec/4_conclusion.tex
\section{Conclusion}

In this demo project, we have introduced a method for efficiently customizing personalized digital avatars, while also reducing 3D modeling costs in certain workflows. We hope that this work will contribute to the widespread adoption of the metaverse, providing users with more engaging experiences.